\documentclass[prd,preprint,nofootinbib,superscriptaddress]{revtex4}

\usepackage{graphicx}
\usepackage{amssymb}
\usepackage{amsmath}
\usepackage{color}

\begin{document}
%%%%%%%%%default command%%%%%%%%%%%
\renewcommand{\thefootnote}{\#\arabic{footnote}}
\newcommand{\rem}[1]{{\bf [#1]}}
\newcommand{\gsim}{ \mathop{}_ {\textstyle \sim}^{\textstyle >} }
\newcommand{\lsim}{ \mathop{}_ {\textstyle \sim}^{\textstyle <} }
\newcommand{\vev}[1]{ \left\langle {#1}  \right\rangle }
\newcommand{\bear}{\begin{array}}  
\newcommand {\eear}{\end{array}}
\newcommand{\bea}{\begin{eqnarray}}   
\newcommand{\eea}{\end{eqnarray}}
\newcommand{\beq}{\begin{equation}}   
\newcommand{\eeq}{\end{equation}}
\newcommand{\bef}{\begin{figure}}  
\newcommand {\eef}{\end{figure}}
\newcommand{\bec}{\begin{center}} 
\newcommand {\eec}{\end{center}}
\newcommand{\non}{\nonumber}  
\newcommand {\eqn}[1]{\beq {#1}\eeq}
\newcommand{\la}{\left\langle}  
\newcommand{\ra}{\right\rangle}
\newcommand{\ds}{\displaystyle}
\newcommand{\red}{\textcolor{red}}
\def\SEC#1{Sec.~\ref{#1}}
\def\FIG#1{Fig.~\ref{#1}}
\def\EQ#1{Eq.~(\ref{#1})}
\def\EQS#1{Eqs.~(\ref{#1})}
\def\lrf#1#2{ \left(\frac{#1}{#2}\right)}
\def\lrfp#1#2#3{ \left(\frac{#1}{#2} \right)^{#3}}
\def\GEV#1{10^{#1}{\rm\,GeV}}
\def\MEV#1{10^{#1}{\rm\,MeV}}
\def\KEV#1{10^{#1}{\rm\,keV}}
\def\REF#1{(\ref{#1})}
\def\lrf#1#2{ \left(\frac{#1}{#2}\right)}
\def\lrfp#1#2#3{ \left(\frac{#1}{#2} \right)^{#3}}
\def\OG#1{ {\cal O}(#1){\rm\,GeV}}
%%%%%%%%%%%%%%%%%%%%%%%%%%%%%

\newcommand{\ah}{A_H}

\title{Leptophilic Dark Matter and AMS-02 Cosmic-ray Positron Flux}

\author{Qing-Hong Cao}
\email{qinghongcao@pku.edu.cn}
\affiliation{Department of Physics and State Key Laboratory of Nuclear Physics
and Technology, Peking University, Beijing 100871, China}
\affiliation{Collaborative Innovation Center of Quantum Matter, Beijing, China}
\affiliation{Center for High Energy Physics, Peking University, Beijing 100871, China}

\author{Chuan-Ren Chen}
\email{crchen@ntnu.edu.tw}
\affiliation{Department of Physics, National Taiwan Normal University, Taipei 116, Taiwan}

\author{Ti Gong}
\email{ ttigong@pku.edu.cn}
\affiliation{Department of Physics and State Key Laboratory of Nuclear Physics
and Technology, Peking University, Beijing 100871, China}

\begin{abstract}
\vspace*{0.5cm}
With the measurement of positron flux published recently by AMS-02 collaboration, we show how the leptophilic dark matter fits the observation.  
We obtain the percentages of different products of dark matter annihilation that can best describe the flux of high energy positrons observed by AMS.  We show that dark matter annihilates predominantly into $\tau\tau$ pair, while both  $ee$ and $\mu\mu$ final states should be less than $20\%$. When gauge boson final states are  included, the best branching ratio of needed $\tau\tau$ mode reduces.   
\end{abstract}

\maketitle

%%%%%%%%%%%%%%%%%%%%%%%%%%%%%
\section{introduction}
\label{sec:1} 
%%%%%%%%%%%%%%%%%%%%%%%%%%%%%
The existence of  dark matter has been firmly  supported by many astrophysical and cosmological observations. However, we have very little knowledge about what  dark matter is made of and how it interacts with particles in the Standard Model. 
Currently, there are many experiments searching for dark matter. The so-called indirect search looks for the products of dark matter annihilation in our Universe. It drew lots of attention recently since both PAMELA~\cite{Adriani:2008zr} and AMS-02~\cite{Aguilar:2013qda,Accardo:2014lma} experiments observed an anomaly in cosmic-ray positron fraction measurement. An excess of positron in the energy region $E\gtrsim 10$ GeV can not be explained by the known sources. It indicates that there exists  a source of primary positrons nearby. 
%An excess of positron in the energy region $E\gtrsim 10$ GeV that can not be explained by the known sources indicates that there exists  a source of primary positrons nearby. 
One possibility for the origin of these high energy positrons is dark matter~\cite{Lin:2014vja,Yuan:2014pka,Jin:2013nta,Yuan:2013eja,Ibe:2013nka,Ibe:2013jya,Dev:2013hka,Ibarra:2013zia,Cirelli:2008pk,Ibe:2014qya}. 

Furthermore, the flux of positron observed by PAMELA shows that the energy spectrum of  positron is harder than background expectation~\cite{Adriani:2013uda,Delahaye:2010ji} for energy of positron larger than about $30$ GeV. This behavior is confirmed by the latest AMS-02 result~\cite{Aguilar:2014mma}.   In addition to cosmic-ray positron data, the antiproton data observed by PAMELA~\cite{Adriani:2010rc} shows that the antiproton flux is consistent with the background, which suggests that the cosmic-ray antiprotons are mainly secondary. Combining cosmic-ray positron and antiproton data, a leptophilic dark matter candidate that annihilates predominately into leptons is attractive since it produces large amount of energetic positrons while the antiproton flux is suppressed. There are several leptophilic dark matter candidates have been studied in the literature~\cite{Chen:2008dh,Yin:2008bs,Fox:2008kb,Bi:2009md,Cao:2009yy}. In this paper, we focus on the leptophilic dark matter scenario in which charged leptons are predominantly produced  when dark matter annihilates. With the precision measurements of positron flux by AMS-02~~\cite{Aguilar:2014mma}, we vary the fraction of individual charged lepton mode ($e^+e^-$, $\mu^+\mu^-$, $\tau^+\tau^-$) and find the best fit for the shape of spectrum. A proper total annihilation cross section is chosen to reach the minimum $\chi^2$.  
For comparison, we also consider the case that the leptophilic dark matter annihilates into massive gauge bosons $W^+W^-$ or $ZZ$. 

The rest of paper is organized as follows. In section~\ref{sec:propagation}, we briefly review the propagation of positron and antiproton, and present the relative parameters in our numerical study. The section~\ref{sec:fitting} is the numerical results showing our fitting with latest AMS-02 result. Finally, we  conclude in section~\ref{sec:conclusion}.

%%%%%%%%%%%%%%%%%%%%%%%%%%%%%
\section{Cosmic-ray positron and antiproton propagation}
\label{sec:propagation} 
%%%%%%%%%%%%%%%%%%%%%%%%%%%%%
%%%%% positron %%%%%

After being produced in the processes of dark matter annihilation, the stable particles will propagate
in the magnetic field of the Milky Way. 
%The motion of the positron can be described by a diffusion equation.
For positrons, the energy spectrum can be obtained by solving the flowing diffusion equation when one consider only the dominant contributions from spatial diffusion and energy lost~\cite{Ibarra:2008qg}, %%
\begin{equation}
\bigtriangledown\cdot\left[K(E,\vec{r})\bigtriangledown f_{e^{+}}\right]+\frac{\partial}{\partial E}\left[b(E,\vec{r})f_{e^{+}}\right]+Q(E,\vec{r})=0,\,\label{eq:e_prop}\end{equation}
where $f_{e^{+}}$ is the number density of $e^{+}$ per unit kinetic
energy, $K(E,\vec{r})$ is the diffusion coefficient, $b(E,\vec{r})$
is the rate of energy loss and $Q(E,\vec{r})$ is the source of 
$e^{+}$. Here the primary positrons are produced in dark matter annihilation,
\[
Q(E,\vec{r})=\frac{1}{2}\left(\frac{\rho(\vec{r})}{m_{DM}}\right)^2\sum_i\left<\sigma v\right>_i\left(\frac{dN_{e^{+}}}{dE}\right)_i,
\]
 where $\rho(\vec{r})$ is the dark matter profile in the Milky Way, $m_{DM}$ is the mass of dark matter, $(dN_{e^{+}}/dE)_i$ is the energy spectrum of $e^{+}$ from dark matter annihilation into 
any possible final state $i$ that generates electrons, with annihilation cross section $\left<\sigma v\right>_i$. 
In our study, $i=e^+e^-,~\mu^+\mu^-,~\tau^+\tau^-$,  $W^+W^-$ or $ZZ$.
The flux of $e^{+}$ originated from dark matter is then given by 
\[
\Phi_{e^{+}}^{DM}(E)=\frac{c}{4\pi}f_{e^{+}}(E,r_\odot),\,
\]
where $c$ is the speed of light and $r_\odot \sim 8.5$ kpc is the distance from the Milky Way center to the Sun. 
We use micrOMEGAs 4~\cite{Belanger:2014vza} with its default settings to calculate the propagation of positrons that originate from dark matter annihilation. The Zhao dark matter profile \cite{Zhao:1995cp} is used.

 In addition to $e^{+}$ flux from dark matter decay, there exists
a secondary $e^{+}$ flux from interactions between cosmic rays and
nuclei in the interstellar medium. 
%% do not need electron flux %%%
%Therefore, the positron fraction is
%\begin{equation}
%\frac{\Phi_{e^{+}}}{\Phi_{e^{+}}+\Phi_{e^{-}}},\,\label{eq:ep_exp}
%\end{equation}
%where $\Phi_{e^+}=\Phi^{DM}_{e^+}+\Phi^{sec}_{e^+}$ and $\Phi_{e^-}=\Phi^{DM}_{e^-}+\Phi^{prim}_{e^-}+\Phi^{sec}_{e^-}$ with the background fluxes of primary electron  $\Phi^{prim}_{e^-}$, secondary electron  $\Phi^{sec}_{e^-}$ and secondary positron  $\Phi^{sec}_{e^+}$ being well approximated as~\cite{Moskalenko:1997gh,Baltz:1998xv} 
%%
The secondary positron  $\Phi^{sec}_{e^+}$ can be well approximated as~\cite{Moskalenko:1997gh,Baltz:1998xv} 
\begin{eqnarray}
%\Phi_{e^{-}}^{prim}(E) & = &\frac{0.16E^{-1.1}}{1+11E^{0.9}+3.2E^{2.15}}\quad{\rm GeV}^{-1}{\rm cm}^{-2}{\rm sec}^{-1}{\rm sr}^{-1},\nonumber \\
%\Phi_{e^{-}}^{sec}(E) & =&\frac{0.7E^{0.7}}{1+110E^{1.5}+600E^{2.9}+580E^{4.2}}\quad{\rm GeV}^{-1}{\rm cm}^{-2}{\rm sec}^{-1}{\rm sr}^{-1},\nonumber \\
\Phi_{e^{+}}^{sec} (E)&=&\frac{4.5E^{0.7}}{1+650E^{2.3}+1500E^{4.2}}\quad~[{\rm GeV}^{-1}{\rm cm}^{-2}{\rm sec}^{-1}{\rm sr}^{-1}],\,\label{eq:bg_e}
\end{eqnarray}
where $E$ is in unit of GeV. 
%We take electron flux from dark matter annihilation $\Phi^{DM}_{e^-} =\Phi^{DM}_{e^+} $.

%%% antiproton %%%
The propagation of antiprotons, neglecting the energy lost, can be described as 
\begin{equation}
K_p\bigtriangledown^2 f_{\bar{p}}(T,\vec{r})- V_c\frac{\partial}{\partial z}f_{\bar{p}}(T,\vec{r})-2h\delta(z)\Gamma_{ann} f_{\bar{p}}(T,\vec{r})+Q(T,\vec{r})=0,\,
\label{eq:diff_ap}
\end{equation}
where $f_{\bar{p}}(T,\vec{r})$ is the number density of antiproton per unit energy, $T$ is the kinetic energy of antiproton, $K_p$ is the diffusion parameter and  $V_c$  is related to the convective wind that tends to push antiprotons away from the Galactic plane and is assumed to be a constant. The third term on the left-hand side of Eq.~(\ref{eq:diff_ap})  represents the annihilation of $\bar{p}$ with the interstellar proton in the Galactic plane where $h$ is the half-height of plane and $\Gamma_{ann}$ is the antiproton annihilation rate. The solution of the interstellar flux of antiproton in the vicinity of solar system is \cite{Cirelli:2008id}
\bea
\Phi^{\rm{IS}}(T,\vec{r}_\odot) & =&\frac{v_{\bar{p}}}{4\pi}f_{\bar{p}}(T,\vec{r}_\odot).                  
\eea
However, when the effects of solar modulation, which are important for the low energy antiprotons, are taken into account, the flux of antiproton obtained at the Earth will be given as \cite{ap:solar1,ap:solar2}
\bea
\frac{d\Phi^\odot_{\bar p}}{d T_\odot} = \frac{p^2_\odot}{p^2_{IS}} \frac{d}{ dT_{IS}}\Phi^{\rm{IS}}(T_{IS},\vec{r}_\odot),
\eea
 where $T_\odot=T_{\rm{IS}}-\phi_F$ is the kinetic energy of antiproton observed at the Earth with $\phi_F$ being the solar modular parameter; $p_\odot $ and $p_{IS}$ are the momentum of antiproton at the Earth and in the interstellar medium, respectively. 
The astrophysical antiproton $\Phi_{\bar{p}}^{bg} $  background can be written as a simple fitting function provided in Ref.~\cite{Nezri:2009jd},
\bea
\Phi_{\bar{p}}^{bg} &=& \frac{0.9t^{-0.9}}{14+30t^{-1.85}+0.08t^{2.3}}\,[\rm{GeV}^{-1}\rm{m}^{-2}\rm{s}^{-1}\rm{sr}^{-1}],
%\Phi_p^{bg} &=& 10^4\frac{0.9t^{-1}}{8+1.1t^{-1.85}+0.8t^{1.68}}\,[\rm{GeV}^{-1}\rm{m}^{-2}\rm{s}^{-1}\rm{sr}^{-1}].
\label{eq:ap_bg}
\eea
where $t$ is the kinetic energy of $\bar p$.
Note that all the propagation parameters are chosen as the default settings in micrOMEGAs. 

%%%%%%%%%%%

%%%%%%%%%%%%%%%%%%%%%%%%%%%%%
\section{Dark Matter and AMS-02 Positron Flux}
\label{sec:fitting} 
%%%%%%%%%%%%%%%%%%%%%%%%%%%%%

\begin{figure}[]
\begin{center}
\includegraphics[scale=0.5,clip]{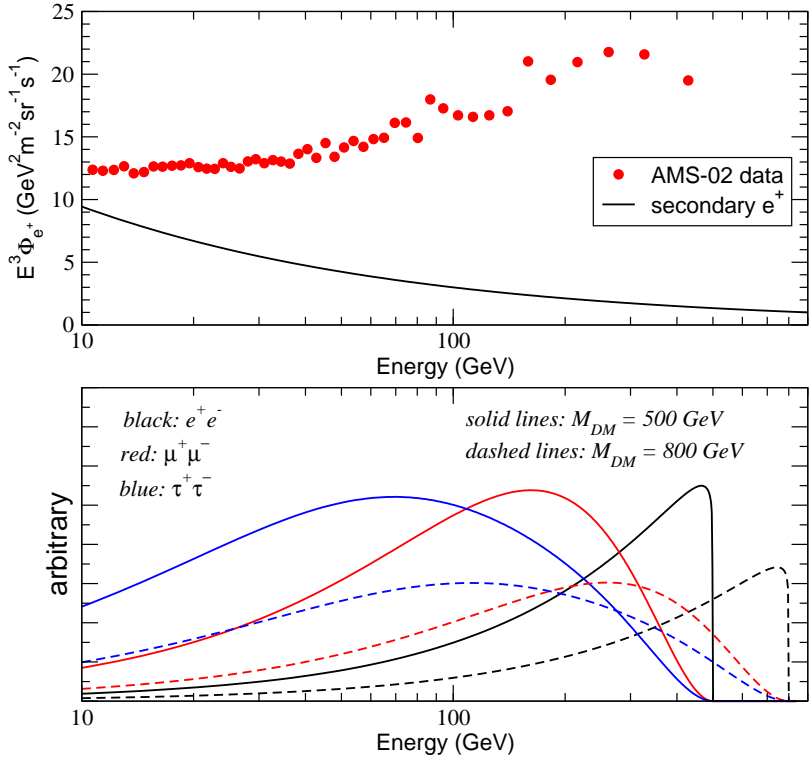}
%\vspace{-1cm}
\caption{(Top) The comparison between AMS-02 updated positron flux data and expected secondary positron flux. (Bottom) The comparison of positron fluxes  from  different final charged leptons in the final state of dark matter annihilation. Note that we multiply different factors to $\mu\mu$ and $\tau\tau$ modes for a better comparison in shape.}
\label{fig:epflux}
\end{center}
\end{figure}

Figure~\ref{fig:epflux} displays the positron flux data observed by AMS-02, compared with the secondary positrons from background. 
We focus only on the high energy positron above 20~GeV such that we do not have to consider the effects of solar modulation that are significant for low energy positrons~\cite{Aguilar:2014mma}.  
%significant for the positron with energy below $10$ GeV~\cite{Aguilar:2014mma}.  
The rise of AMS-02 data at $E\sim 30$ GeV shows that the spectral index is larger than $-3$, and such a behavior certainly can not be explained by the secondary positrons.
In the bottom panel of Fig.~\ref{fig:epflux}, we plot the shapes of positron flux ($E^3\Phi_{e^+}$) for different leptonic final states after propagation. The behavior of cosmic-ray positron varies significantly for different lepton modes. 
The $ee$ mode, as originating directly from dark matter annihilation, yields the hardest cosmic-ray positron. The flux peaks at energy around the mass of dark matter and drops quickly.   The $\mu\mu$ mode peaks at energy that is close to half of dark matter mass, while the $\tau\tau$ mode reaches the top at energy even lower.  

\begin{figure}[]
\begin{center}
\includegraphics[scale=0.4,clip]{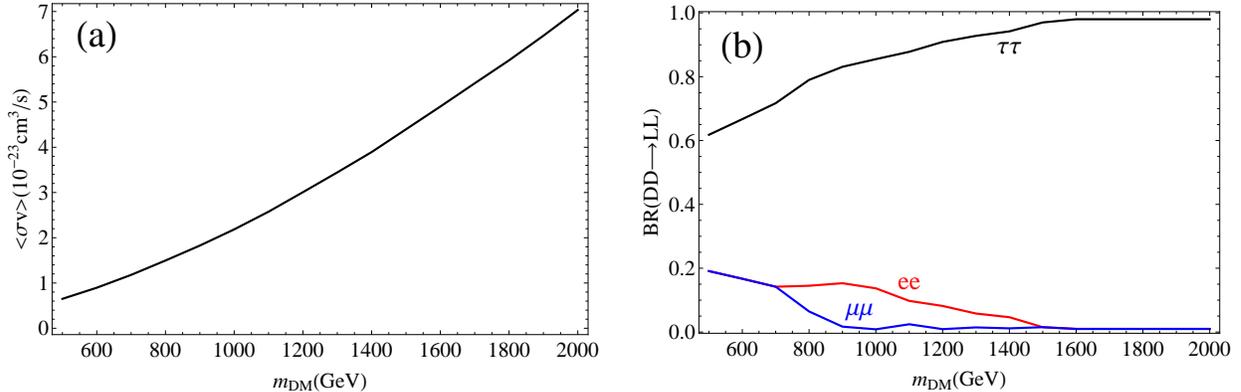}
\caption{(a) The annihilation cross section needed for the best fit of AMS-02 updated positron flux data. (b) The fractions of leptonic modes of dark matter annihilation to fit AMS-02 updated data. }
\label{fit1}
\end{center}
\end{figure}
To explain the AMS-02 positron flux with leptophilic dark matter,
we adopt a minimal $\chi^2$-fit to obtain the best values of the total annihilation cross section as well as the  ratios of the leptonic modes from dark matter annihilation for a chosen dark matter mass; see Fig.~\ref{fit1}. 
Among all the AMS-02 data, we consider the positron with energy larger than 20~GeV to safely avoid the effects of solar modulation in our $\chi^2$-fit. 
Figure~\ref{fit1}(a) shows the annihilation cross section is around $10^{-23}-10^{-22}{\rm cm}^{3}/{\rm s}$, which is about 1000 times larger than the thermal cross section $\left<\sigma v\right>=3\times 10^{-26}{\rm cm}^3/{\rm s}$. 
We take the mass of  dark matter candidate to be larger than 500~GeV in order to explain the arising feature of positron energy spectrum. 
%If the positron arises from the primary annihilation, its energy exhibits a sharp peak around dark matter mass and is harder than those positrons from $\mu\mu$ and $\tau\tau$ modes. 
In Fig.~\ref{fit1}(b), the $\chi^2$-fit shows that the $ee$-mode should be less than 20\% for a large range of dark matter mass. The $\mu\mu$ mode is also disfavored as the dark matter mass increases; see the blue curve. The $\tau\tau$ mode tends to be dominant for a heavy dark matter candidate.
Our fit shows that the three leptonic modes are more or less democratic for a light dark matter around 500~GeV. For a dark matter heavier than 1~TeV, the $\tau\tau$ mode is well above the other two modes. The dark matter model  should consist of a non-trial flavor structure, e.g. the Higgs portal model. 
In the Higgs portal models the dark matter candidate could predominately decay into vector boson pairs. For comparison, we add in the contributions of gauge boson modes, $W^+W^-$ and $ZZ$, to see how the fitting can be improved. 
\begin{figure}[]
\begin{center}
\includegraphics[scale=0.4,clip]{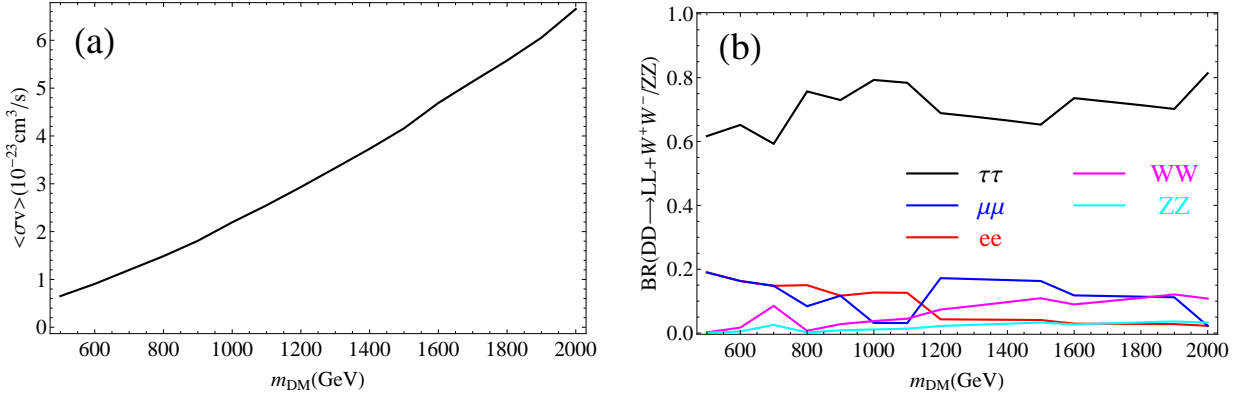}
\caption{(a) The annihilation cross section needed for the  best fit of AMS-02 positron flux data, including $WW$ and $ZZ$ gauge boson modes. (b) The fractions of different final states of dark matter annihilation to fit AMS-02 positron flux data. }
\label{fitvv}
\end{center}
\end{figure}
\begin{figure}[htbp]
\begin{center}
\includegraphics[scale=0.38,clip]{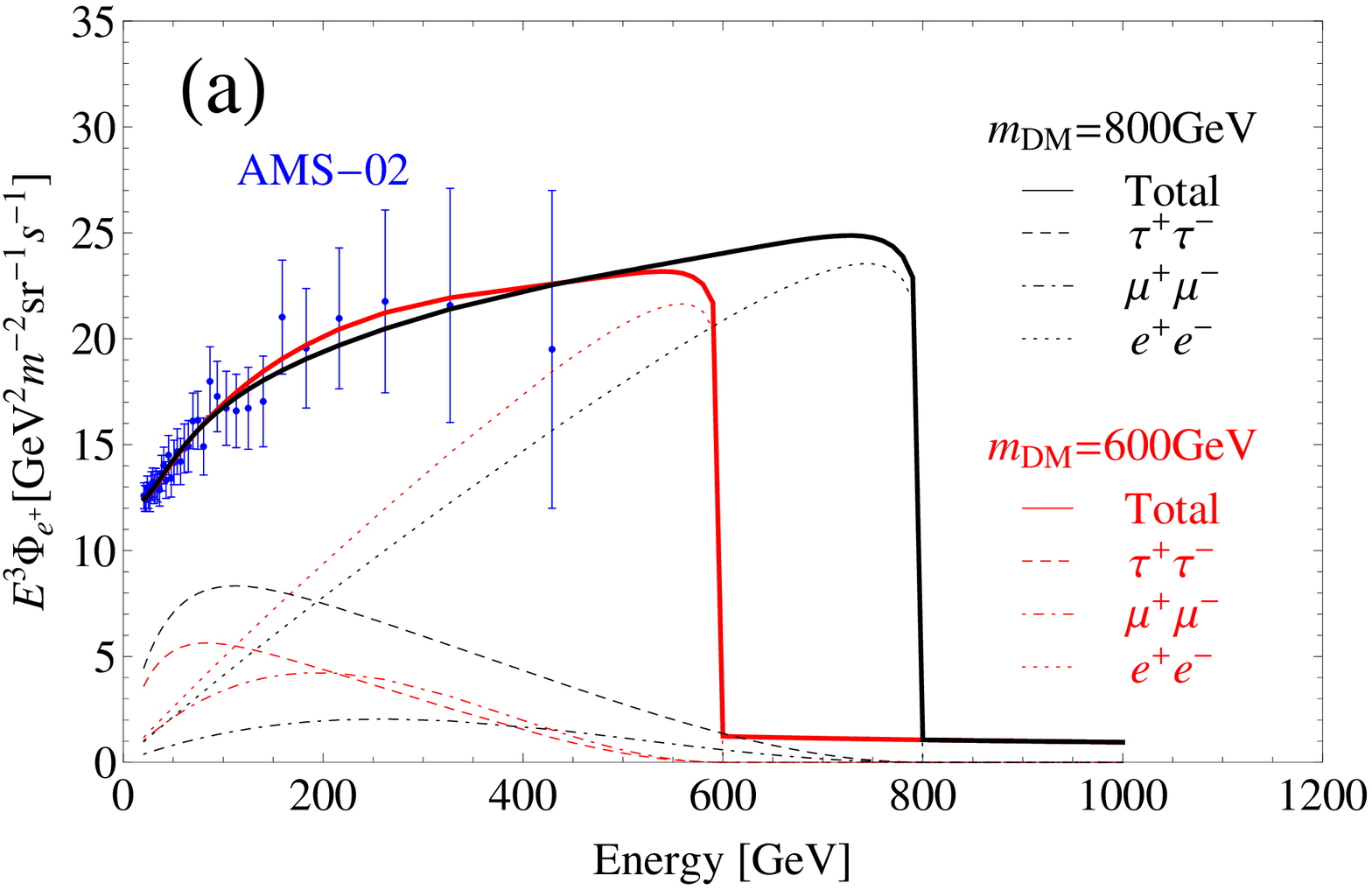}
\includegraphics[scale=0.38,clip]{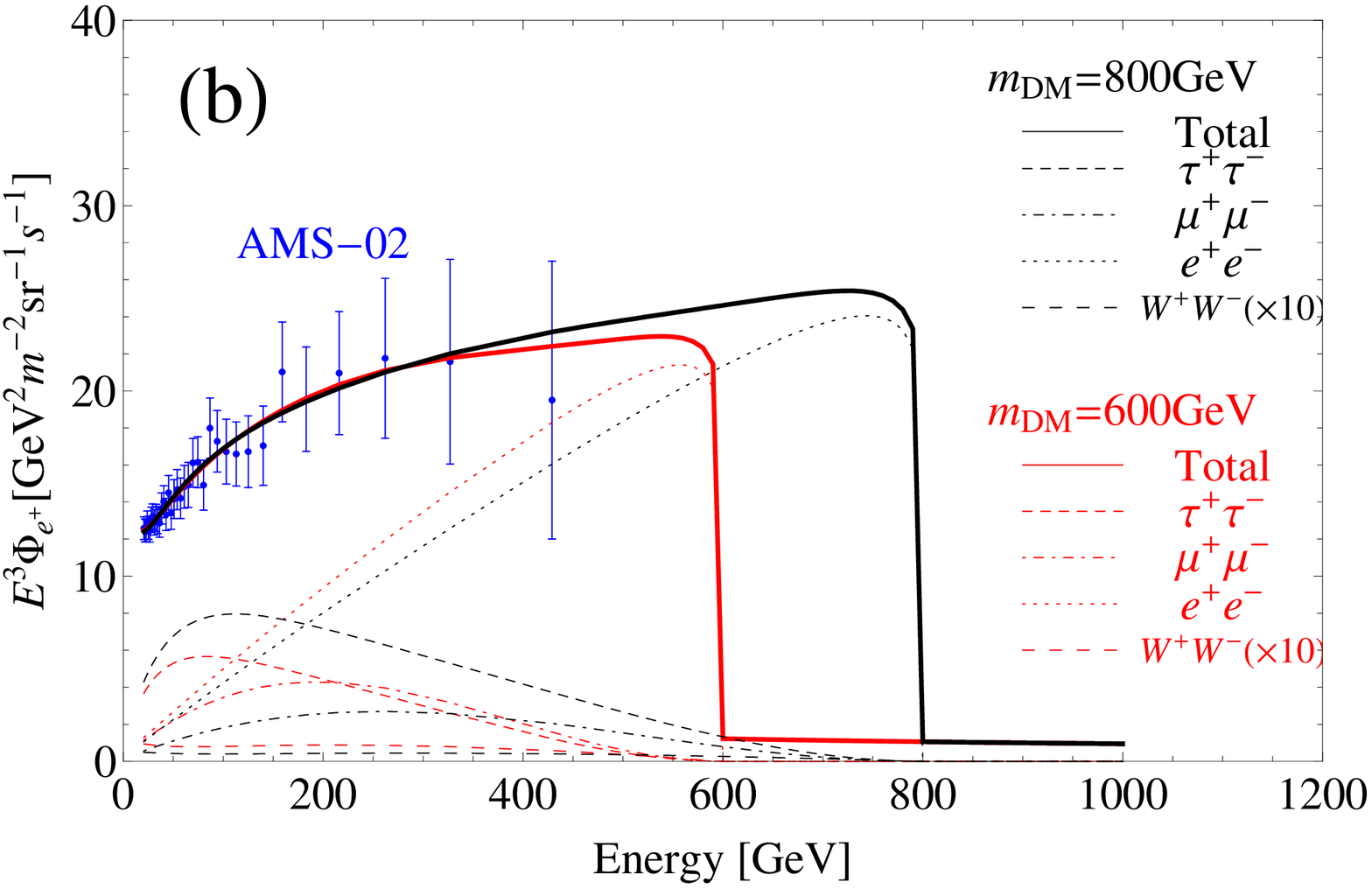}
\caption{(a) Comparison between AMS-02 positron flux data~\cite{Aguilar:2014mma} and prediction from dark matter annihilation with best fit of annihilation cross section and fractions of lepton final states.  (b) Same as Left, but including $WW$ and $ZZ$ gauge boson modes. }
\label{epdata}
\end{center}
\end{figure}

Figure~\ref{fitvv} displays the result of our best values of lepton plus gauge boson modes. The needed annihilation cross section is about the same as the case of purely leptonic final states. However, the fraction of individual mode is changed as $m_{DM}$ is larger than about $600$ GeV. While the  $\tau\tau$ mode is still dominant, but it does not exceed  $80\%$. The $WW$ mode plays a significant role for a heavier dark matter and reaches about $10\%$ when $m_{DM}\gtrsim 1.4$ TeV. 
We notice that the $ZZ$ mode is not important in the fitting.
In Fig.~\ref{epdata}, we show the flux of positron predicted by the dark matter annihilation using the best values of annihilation cross section and fractions of different annihilation products. The AMS-02 latest data (shown as the blue points with error bars) is also plotted for comparison. It is obvious that the dark matter could explain the observation well. Here, we only present two benchmark dark matter masses, $600$ GeV and $800$ GeV, for reference. The flux is relatively flat for very high energy positrons and then decreases to pure background prediction at the energy equal to the mass of dark matter. Therefore, the energy where the excess of positron flux vanishes will give us information about the mass of dark matter.

\begin{figure}
\begin{center}
\includegraphics[scale=0.4,clip]{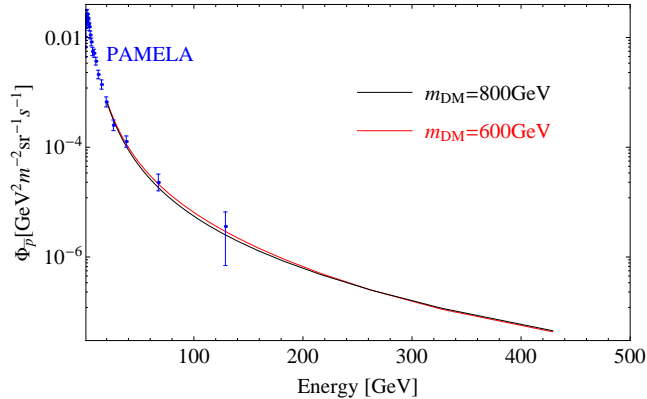}
\caption{The antiproton flux generated by the $WW/ZZ$ modes from dark matter annihilation. The blue points represent the PAMELA data~\cite{Adriani:2010rc}. }
\label{antip}
\end{center}
\end{figure}

Including the $WW$ mode improved the $\chi^2$ fit in general. Among the dark matter masses we study, the overall best $\chi^2$ is obtained when $m_{DM} = 600$ GeV for both pure lepton case and the situation that the $WW$ and $ZZ$ modes are considered. However, the $WW$ final state not only generates positrons, but also produces antiprotons that have been measured to be consistent with the background that mainly contains  secondary antiprotons~\cite{Orito:1999re,Abe:2008sh,Asaoka:2001fv,Boezio:2001ac,Adriani:2008zq,Adriani:2010rc}.   We show in Fig.~\ref{antip} the flux of antiproton predicted by the dark matter annihilation into the $WW$ final state, with the latest PAMELA observation~\cite{Adriani:2010rc}. Antiproton flux from dark matter annihilation agrees with observation quite well for $600$ GeV and $800$ GeV dark matter.

%

%%%%%%%%%%%%%%%%%%%%%%%%%%%%%
\section{Discussion and Conclusions}
\label{sec:conclusion} 
%%%%%%%%%%%%%%%%%%%%%%%%%%%%%

The AMS-02 collaboration recently published the results of cosmic-ray positrons. 
The measurement of positron flux for the energy up to $500$ GeV, with good energy resolution and small uncertainty, allows us to gain more information about dark matter, if we assume dark matter is responsible to the disagreement between data and the known background.
In this paper, we study how the leptophilic dark matter can provide explanation of the AMS-02 positron flux data.  
Our results show that the leptophilic dark matter prefers to annihilate into  $\tau\tau$ final state. 
 Both the $ee$ and $\mu\mu$ modes should be less than $20\%$, and the $\tau\tau$ mode always dominants. The percentage of the $\tau\tau$ mode could reach almost $100\%$ for the dark matter heavier than about $1.5$ TeV unless the $WW$ final state is available. We learn that the $WW$ mode plays a significant role for dark matter heavier than about $1.1$ TeV. The existence of $WW$ final state also brings up the importance of the $\mu\mu$ mode. Combination of the $\mu\mu$ and $WW$ modes could contribute about $30\%$ of total annihilation cross section, while the $ee$ and $ZZ$ modes can be neglected for $m_{DM}\gtrsim 1.2$ TeV.  We also calculate the antiproton flux generated from the $WW$ final state, and the result is consistent with current PAMELA data.

Finally, we are aware that there is a tension between the $\tau\tau$ final state and cosmic $\gamma$-ray data~\cite{Ackermann:2011wa, Cirelli:2009dv}. However, as we have shown that adding the $WW$ mode in the leptophilic dark matter annihilation reduces the needed fraction of the $\tau\tau$ mode for the best fit. Furthermore, although the result is not shown, we realize that the $\tau\tau$ contribution can be further reduced when quark final states are included. Surely, with contributions of the $WW$ and quark modes, additional antiproton flux will be produced. Therefore, the future antiproton results from AMS-02 are highly expected to tell us more properties of dark matter.

\begin{acknowledgments}
The work of QHC and TG is supported in part by the National Science Foundation of China under Grand No. 11275009. The work of CRC is supported in part by the National Science Council under Grant No.~NSC 102-2112-M-003-001-MY3. 
\end{acknowledgments}


\begin{thebibliography}{10}

%\cite{Adriani:2008zr}
\bibitem{Adriani:2008zr} 
  O.~Adriani {\it et al.}  [PAMELA Collaboration],
  %``An anomalous positron abundance in cosmic rays with energies 1.5-100 GeV,''
  Nature {\bf 458}, 607 (2009)
  [arXiv:0810.4995 [astro-ph]].
  %%CITATION = ARXIV:0810.4995;%%
  %1306 citations counted in INSPIRE as of 22 Sep 2014
  
  %\cite{Aguilar:2013qda}
\bibitem{Aguilar:2013qda} 
  M.~Aguilar {\it et al.}  [AMS Collaboration],
  %``First Result from the Alpha Magnetic Spectrometer on the International Space Station: Precision Measurement of the Positron Fraction in Primary Cosmic Rays of 0.5Ð350 GeV,''
  Phys.\ Rev.\ Lett.\  {\bf 110}, 141102 (2013).
  %%CITATION = PRLTA,110,141102;%%
  %248 citations counted in INSPIRE as of 21 Sep 2014

%\cite{Accardo:2014lma}
\bibitem{Accardo:2014lma} 
  L.~Accardo [AMS Collaboration],
  %``High Statistics Measurement of the Positron Fraction in Primary Cosmic Rays of 0.5Ð500 GeV with the Alpha Magnetic Spectrometer on the International Space Station,''
  Phys.\ Rev.\ Lett.\  {\bf 113}, 121101 (2014).
  %%CITATION = PRLTA,113,121101;%%
  
%%%%%%%% dark matter positron 
%\cite{Cirelli:2008pk}
\bibitem{Cirelli:2008pk} 
  M.~Cirelli, M.~Kadastik, M.~Raidal and A.~Strumia,
  %``Model-independent implications of the e+-, anti-proton cosmic ray spectra on properties of Dark Matter,''
  Nucl.\ Phys.\ B {\bf 813}, 1 (2009)
  [Addendum-ibid.\ B {\bf 873}, 530 (2013)]
  [arXiv:0809.2409 [hep-ph]].
  %%CITATION = ARXIV:0809.2409;%%
  %426 citations counted in INSPIRE as of 24 Sep 2014
   

%\cite{Yuan:2013eja}
\bibitem{Yuan:2013eja} 
  Q.~Yuan, X.~J.~Bi, G.~M.~Chen, Y.~Q.~Guo, S.~J.~Lin and X.~Zhang,
  %``Implications of the AMS-02 positron fraction in cosmic rays,''
  Astropart.\ Phys.\  {\bf 60}, 1 (2015)
  [arXiv:1304.1482 [astro-ph.HE]].
  %%CITATION = ARXIV:1304.1482;%%
  %38 citations counted in INSPIRE as of 24 Sep 2014

%\cite{Ibe:2013nka}
\bibitem{Ibe:2013nka} 
  M.~Ibe, S.~Iwamoto, S.~Matsumoto, T.~Moroi and N.~Yokozaki,
  %``Recent Result of the AMS-02 Experiment and Decaying Gravitino Dark Matter in Gauge Mediation,''
  JHEP {\bf 1308}, 029 (2013)
  [arXiv:1304.1483 [hep-ph]].
  %%CITATION = ARXIV:1304.1483;%%
  %19 citations counted in INSPIRE as of 24 Sep 2014

 %\cite{Jin:2013nta}
\bibitem{Jin:2013nta} 
  H.~B.~Jin, Y.~L.~Wu and Y.~F.~Zhou,
  %``Implications of the first AMS-02 measurement for dark matter annihilation and decay,''
  JCAP {\bf 1311}, 026 (2013)
  [arXiv:1304.1997 [hep-ph]].
  %%CITATION = ARXIV:1304.1997;%%
  %37 citations counted in INSPIRE as of 24 Sep 2014


%\cite{Ibe:2013jya}
\bibitem{Ibe:2013jya} 
  M.~Ibe, S.~Matsumoto, S.~Shirai and T.~T.~Yanagida,
  %``AMS-02 Positrons from Decaying Wino in the Pure Gravity Mediation Model,''
  JHEP {\bf 1307}, 063 (2013)
  [arXiv:1305.0084 [hep-ph]].
  %%CITATION = ARXIV:1305.0084;%%
  %18 citations counted in INSPIRE as of 24 Sep 2014

%\cite{Dev:2013hka}
\bibitem{Dev:2013hka}
  P.~S.~B.~Dev, D.~K.~Ghosh, N.~Okada and I.~Saha,
  %``Neutrino Mass and Dark Matter in light of recent AMS-02 results,''
  Phys.\ Rev.\ D {\bf 89} (2014) 095001
  [arXiv:1307.6204 [hep-ph]].
  %%CITATION = ARXIV:1307.6204;%%
  %8 citations counted in INSPIRE as of 21 Sep 2014

%\cite{Ibarra:2013zia}
\bibitem{Ibarra:2013zia} 
  A.~Ibarra, A.~S.~Lamperstorfer and J.~Silk,
  %``Dark matter annihilations and decays after the AMS-02 positron measurements,''
  Phys.\ Rev.\ D {\bf 89}, 063539 (2014)
  [arXiv:1309.2570 [hep-ph]].
  %%CITATION = ARXIV:1309.2570;%%
  %23 citations counted in INSPIRE as of 21 Sep 2014  
 
  
%\cite{Yuan:2014pka}
\bibitem{Yuan:2014pka} 
  Q.~Yuan and X.~J.~Bi,
  %``Systematic study of the uncertainties in fitting the cosmic positron data by AMS-02,''
  arXiv:1408.2424 [astro-ph.HE].
  %%CITATION = ARXIV:1408.2424;%%
  %1 citations counted in INSPIRE as of 24 Sep 2014  

%\cite{Lin:2014vja}
\bibitem{Lin:2014vja} 
  S.~J.~Lin, Q.~Yuan and X.~J.~Bi,
  %``Quantitative study of the AMS-02 electron/positron spectra: implications for the pulsar and dark matter properties,''
  arXiv:1409.6248 [astro-ph.HE].
  %%CITATION = ARXIV:1409.6248;%%
 
  
%\cite{Ibe:2014qya}
\bibitem{Ibe:2014qya} 
  M.~Ibe, S.~Matsumoto, S.~Shirai and T.~T.~Yanagida,
  %``Mass of Decaying Wino from AMS-02 2014,''
  arXiv:1409.6920 [hep-ph].
  %%CITATION = ARXIV:1409.6920;%%
    
%%%%%%%%%%%%%%%%%%%%    


%\cite{Adriani:2013uda}
\bibitem{Adriani:2013uda} 
  O.~Adriani {\it et al.}  [PAMELA Collaboration],
  %``Cosmic-Ray Positron Energy Spectrum Measured by PAMELA,''
  Phys.\ Rev.\ Lett.\  {\bf 111}, no. 8, 081102 (2013)
  [arXiv:1308.0133 [astro-ph.HE]].
  %%CITATION = ARXIV:1308.0133;%%
  %25 citations counted in INSPIRE as of 22 Sep 2014  

%\cite{Delahaye:2010ji}
\bibitem{Delahaye:2010ji}
  T.~Delahaye, J.~Lavalle, R.~Lineros, F.~Donato and N.~Fornengo,
  %``Galactic electrons and positrons at the Earth:new estimate of the primary and secondary fluxes,''
  Astron.\ Astrophys.\  {\bf 524} (2010) A51
  [arXiv:1002.1910 [astro-ph.HE]].
  %%CITATION = ARXIV:1002.1910;%%
  %65 citations counted in INSPIRE as of 22 Sep 2014

  
%\cite{Aguilar:2014mma}
\bibitem{Aguilar:2014mma} 
  M.~Aguilar [AMS Collaboration],
  %``Electron and Positron Fluxes in Primary Cosmic Rays Measured with the Alpha Magnetic Spectrometer on the International Space Station,''
  Phys.\ Rev.\ Lett.\  {\bf 113}, 121102 (2014).
  %%CITATION = PRLTA,113,121102;%%  

%\cite{Adriani:2010rc}
\bibitem{Adriani:2010rc} 
  O.~Adriani {\it et al.}  [PAMELA Collaboration],
  %``PAMELA results on the cosmic-ray antiproton flux from 60 MeV to 180 GeV in kinetic energy,''
  Phys.\ Rev.\ Lett.\  {\bf 105}, 121101 (2010)
  [arXiv:1007.0821 [astro-ph.HE]].
  %%CITATION = ARXIV:1007.0821;%%
  %283 citations counted in INSPIRE as of 23 Sep 2014  
  
%
% Leptophilic dark matter 
%
%\cite{Chen:2008dh}
\bibitem{Chen:2008dh} 
  C.~R.~Chen and F.~Takahashi,
  %``Cosmic rays from Leptonic Dark Matter,''
  JCAP {\bf 0902}, 004 (2009)
  [arXiv:0810.4110 [hep-ph]].
  %%CITATION = ARXIV:0810.4110;%%
  %91 citations counted in INSPIRE as of 24 Sep 2014

%\cite{Yin:2008bs}
\bibitem{Yin:2008bs} 
  P.~f.~Yin, Q.~Yuan, J.~Liu, J.~Zhang, X.~j.~Bi and S.~h.~Zhu,
  %``PAMELA data and leptonically decaying dark matter,''
  Phys.\ Rev.\ D {\bf 79}, 023512 (2009)
  [arXiv:0811.0176 [hep-ph]].
  %%CITATION = ARXIV:0811.0176;%%
  %163 citations counted in INSPIRE as of 24 Sep 2014

%\cite{Fox:2008kb}
\bibitem{Fox:2008kb} 
  P.~J.~Fox and E.~Poppitz,
  %``Leptophilic Dark Matter,''
  Phys.\ Rev.\ D {\bf 79}, 083528 (2009)
  [arXiv:0811.0399 [hep-ph]].
  %%CITATION = ARXIV:0811.0399;%%
  %193 citations counted in INSPIRE as of 24 Sep 2014

%\cite{Bi:2009md}
\bibitem{Bi:2009md} 
  X.~J.~Bi, P.~H.~Gu, T.~Li and X.~Zhang,
  %``ATIC and PAMELA Results on Cosmic e+- Excesses and Neutrino Masses,''
  JHEP {\bf 0904}, 103 (2009)
  [arXiv:0901.0176 [hep-ph]].
  %%CITATION = ARXIV:0901.0176;%%
  %37 citations counted in INSPIRE as of 24 Sep 2014
  
%\cite{Cao:2009yy}
\bibitem{Cao:2009yy} 
  Q.~H.~Cao, E.~Ma and G.~Shaughnessy,
  %``Dark Matter: The Leptonic Connection,''
  Phys.\ Lett.\ B {\bf 673}, 152 (2009)
  [arXiv:0901.1334 [hep-ph]].
  %%CITATION = ARXIV:0901.1334;%%
  %64 citations counted in INSPIRE as of 24 Sep 2014


%\cite{Ibarra:2008qg}
\bibitem{Ibarra:2008qg} 
  A.~Ibarra and D.~Tran,
  %``Antimatter Signatures of Gravitino Dark Matter Decay,''
  JCAP {\bf 0807}, 002 (2008)
  [arXiv:0804.4596 [astro-ph]].
  %%CITATION = ARXIV:0804.4596;%%
  %106 citations counted in INSPIRE as of 21 Sep 2014

%\cite{Belanger:2014vza}
\bibitem{Belanger:2014vza} 
  G.~Belanger, F.~Boudjema, A.~Pukhov and A.~Semenov,
  %``micrOMEGAs4.1: two dark matter candidates,''
  arXiv:1407.6129 [hep-ph].
  %%CITATION = ARXIV:1407.6129;%%


%\cite{Moskalenko:1997gh}
\bibitem{Moskalenko:1997gh} 
  I.~V.~Moskalenko and A.~W.~Strong,
  %``Production and propagation of cosmic ray positrons and electrons,''
  Astrophys.\ J.\  {\bf 493}, 694 (1998)
  [astro-ph/9710124].
  %%CITATION = ASTRO-PH/9710124;%%
  %393 citations counted in INSPIRE as of 21 Sep 2014
  
%\cite{Baltz:1998xv}
\bibitem{Baltz:1998xv} 
  E.~A.~Baltz and J.~Edsjo,
  %``Positron propagation and fluxes from neutralino annihilation in the halo,''
  Phys.\ Rev.\ D {\bf 59}, 023511 (1998)
  [astro-ph/9808243].
  %%CITATION = ASTRO-PH/9808243;%%
  %341 citations counted in INSPIRE as of 21 Sep 2014

%\cite{Cirelli:2008id}
\bibitem{Cirelli:2008id}
  M.~Cirelli, R.~Franceschini and A.~Strumia,
  %``Minimal Dark Matter predictions for galactic positrons, anti-protons,
  %photons,''
  Nucl.\ Phys.\  B {\bf 800}, 204 (2008)
  [arXiv:0802.3378 [hep-ph]].
  %%CITATION = NUPHA,B800,204;%%

%\cite{ap:solar1}
\bibitem{ap:solar1}
  L.J.~Gleeson and W. I. ~Axford,
  Astrophys.\ J.\  {\bf 149}, L115 (1967);
%
Astrophys.\ J.\  {\bf 154}, 1011 (1968)

%\cite{ap:solar2}
\bibitem{ap:solar2}
  J. S.~Perko,
  Astron. Astrophys.\ J.\  {\bf 184}, 119 (1987)

%\cite{Nezri:2009jd}
\bibitem{Nezri:2009jd}
  E.~Nezri, M.~H.~G.~Tytgat and G.~Vertongen,
  %``Positrons and antiprotons from inert doublet model dark matter,''
  arXiv:0901.2556 [hep-ph].
  %%CITATION = ARXIV:0901.2556;%%


%\cite{Zhao:1995cp}
\bibitem{Zhao:1995cp} 
  H.~Zhao,
  %``Analytical models for galactic nuclei,''
  Mon.\ Not.\ Roy.\ Astron.\ Soc.\  {\bf 278}, 488 (1996)
  [astro-ph/9509122].
  %%CITATION = ASTRO-PH/9509122;%%
  %214 citations counted in INSPIRE as of 22 Sep 2014

% BESS 95+97
%\cite{Orito:1999re}
\bibitem{Orito:1999re}
  S.ï¿œàOrito {\it et al.}  [BESS Collaboration],
  %``Precision measurement of cosmic-ray antiproton spectrum,''
  Phys.\ Rev.\ Lett.\  {\bf 84}, 1078 (2000)
  [arXiv:astro-ph/9906426].
  %%CITATION = PRLTA,84,1078;%%


% BESS-Polar
%\cite{Abe:2008sh}
\bibitem{Abe:2008sh}
  K.ï¿œàAbe {\it et al.},
  %``Measurement of cosmic-ray low-energy antiproton spectrum with the first
  %BESS-Polar Antarctic flight,''
  Phys.\ Lett.\  B {\bf 670}, 103 (2008)
  [arXiv:0805.1754 [astro-ph]].
  %%CITATION = PHLTA,B670,103;%%
  

% BESS 1999 2000
%\cite{Asaoka:2001fv}
\bibitem{Asaoka:2001fv}
  Y.ï¿œàAsaoka {\it et al.},
  %``Measurements of cosmic-ray low-energy antiproton and proton spectra in  a
  %transient period of the solar field reversal,''
  Phys.\ Rev.\ Lett.\  {\bf 88}, 051101 (2002)
  [arXiv:astro-ph/0109007].
  %%CITATION = PRLTA,88,051101;%%
  

% CAPRICE 98
%\cite{Boezio:2001ac}
\bibitem{Boezio:2001ac}
  M.ï¿œàBoezio {\it et al.}  [WiZard/CAPRICE Collaboration],
  %``The cosmic-ray anti-proton flux between 3-GeV and 49-GeV,''
  Astrophys.\ J.\  {\bf 561}, 787 (2001)
  [arXiv:astro-ph/0103513].
  %%CITATION = ASJOA,561,787;%%  
  
%\cite{Adriani:2008zq}
\bibitem{Adriani:2008zq} 
  O.~Adriani, G.~C.~Barbarino, G.~A.~Bazilevskaya, R.~Bellotti, M.~Boezio, E.~A.~Bogomolov, L.~Bonechi and M.~Bongi {\it et al.},
  %``A new measurement of the antiproton-to-proton flux ratio up to 100 GeV in the cosmic radiation,''
  Phys.\ Rev.\ Lett.\  {\bf 102}, 051101 (2009)
  [arXiv:0810.4994 [astro-ph]].
  %%CITATION = ARXIV:0810.4994;%%
  %455 citations counted in INSPIRE as of 23 Sep 2014
  

%\cite{Ackermann:2011wa}
\bibitem{Ackermann:2011wa} 
  M.~Ackermann {\it et al.}  [Fermi-LAT Collaboration],
  %``Constraining Dark Matter Models from a Combined Analysis of Milky Way Satellites with the Fermi Large Area Telescope,''
  Phys.\ Rev.\ Lett.\  {\bf 107}, 241302 (2011)
  [arXiv:1108.3546 [astro-ph.HE]];
  %%CITATION = ARXIV:1108.3546;%%
  %352 citations counted in INSPIRE as of 24 Sep 2014
%\cite{Ackermann:2012rg}
%\bibitem{Ackermann:2012rg} 
  M.~Ackermann {\it et al.}  [LAT Collaboration],
  %``Constraints on the Galactic Halo Dark Matter from Fermi-LAT Diffuse Measurements,''
  Astrophys.\ J.\  {\bf 761}, 91 (2012)
  [arXiv:1205.6474 [astro-ph.CO]].
  %%CITATION = ARXIV:1205.6474;%%
  %98 citations counted in INSPIRE as of 24 Sep 2014

 %\cite{Cirelli:2009dv}
\bibitem{Cirelli:2009dv} 
  M.~Cirelli, P.~Panci and P.~D.~Serpico,
  %``Diffuse gamma ray constraints on annihilating or decaying Dark Matter after Fermi,''
  Nucl.\ Phys.\ B {\bf 840}, 284 (2010)
  [arXiv:0912.0663 [astro-ph.CO]];
  %%CITATION = ARXIV:0912.0663;%%
  %141 citations counted in INSPIRE as of 24 Sep 2014 
%\cite{Papucci:2009gd}
%\bibitem{Papucci:2009gd} 
  M.~Papucci and A.~Strumia,
  %``Robust implications on Dark Matter from the first FERMI sky gamma map,''
  JCAP {\bf 1003}, 014 (2010)
  [arXiv:0912.0742 [hep-ph]].
  %%CITATION = ARXIV:0912.0742;%%
  %99 citations counted in INSPIRE as of 24 Sep 2014     
  
     
\end{thebibliography}
\end{document}